**Measurement of the motional sidebands of a nanogram-scale oscillator in the quantum regime**

M. Underwood,[1] D. Mason,[1] D. Lee,[1,2] H. Xu,[1] L. Jiang,[1] A. B. Shkarin,[1]
K. Børkje,[1,3] S. M. Girvin,[1,4] and J. G. E. Harris[1,4]

[1] Department of Physics, Yale University, New Haven, CT, 06511, USA
[2] Department of Physics, University of California, Santa Barbara, CA, 93106, USA
[3] Royal Norwegian Naval Academy, Bergen, Norway
[4] Department of Applied Physics, Yale University, New Haven, CT, 06511, USA

We describe measurements of the motional sidebands produced by a mechanical oscillator (with effective mass 43 ng and resonant frequency 705 kHz) that is placed in an optical cavity and cooled close to its quantum ground state. The red and blue sidebands (corresponding to Stokes and anti-Stokes scattering) from a single laser beam are recorded simultaneously via a heterodyne measurement. The oscillator's mean phonon number $\bar{n}$ is inferred from the ratio of the sidebands, and reaches a minimum value of 0.84 ± 0.22 (corresponding to a mode temperature $T = 28 ± 7$ μK). We also infer $\bar{n}$ from the calibrated area of each of the two sidebands, and from the oscillator's total damping. The values of $\bar{n}$ inferred from these four methods are in close agreement. The behavior of the sidebands as a function of the oscillator's temperature agrees well with theory that includes the quantum fluctuations of both the cavity field and the mechanical oscillator.

Cavity optomechanical systems operating in the quantum regime are expected to play an important role in interfacing disparate quantum systems, advancing the coherent control of electromagnetic fields and mechanical oscillators, detecting astrophysical gravitational waves, constraining modifications to orthodox quantum mechanics, and testing hypotheses about quantum gravity.[1,2,3,4,5,6,7,8,9,10,11] The utility of optomechanical systems in these areas (and others) reflects the particular combination of features they offer: long relaxation times, unitary coupling to electromagnetic fields across frequency domains from the microwave to the visible, and access to the quantum behavior of massive objects.

Optomechanical experiments have been based primarily on systems in which the mechanical oscillator and the cavity field are prepared in Gaussian states, couple weakly to each other at the quantum level (i.e., the bare optomechanical coupling rate $g_0$ is much less than the oscillator frequency $\omega_m$ and the cavity damping rate $\kappa$), and are probed via linear measurements of the electromagnetic field leaving the cavity. (Some optomechanics experiments have demonstrated nonlinear measurements of the cavity fields,[12,13] although without resolving non-Gaussian behavior.) Within this paradigm of Gaussian states, weak coupling, and linear measurements, quantum effects can manifest themselves in the apparent fluctuation of quantities which, according to classical mechanics, could be noiseless.[14] Depending on the specific type of measurement, these quantum fluctuations may be ascribed to the cavity field, the mechanical oscillator, or both.[15,16]

One example of such an experiment is a heterodyne measurement of the light leaving an optomechanical cavity that is driven on resonance by a single laser. In a classical description of this experiment, thermal motion of the mechanical oscillator inside the cavity adds modulation sidebands to the laser beam. In the spectrum of the heterodyne signal, the area of these sidebands will be equal, and will be proportional to the oscillator's temperature.

In the quantum treatment described in Refs.[15,16] of the same measurement, the heterodyne spectrum arises from four distinct components: (*i*) the quantum fluctuations of the electromagnetic field, which results in a noise floor equivalent to shot noise; (*ii*) the oscillator's thermal motion, which produces sidebands each with area proportional to the oscillator's mean phonon number $\bar{n}$ (as in the classical case described above); (*iii*) the oscillator's zero point motion, which makes an additional contribution to each sideband that is equivalent to increasing $\bar{n}$ by ½; and (*iv*) the oscillator's response to the quantum fluctuations of the cavity field, which makes a contribution to the Stokes (red) sideband that is equivalent to increasing $\bar{n}$ by ½ and a contribution to the anti-Stokes (blue) sideband that is equivalent to decreasing $\bar{n}$ by ½. The sign difference between the two contributions from (*iv*) reflects the presence of correlations between the quantum fluctuations of the electromagnetic field and the oscillator's motion.

These four components of the heterodyne spectrum are illustrated schematically in Fig. 1(a). Although a measurement of the heterodyne signal only reveals the sum of these contributions, a complete explanation of the full spectrum (particularly as a function of the oscillator's temperature) requires all four components. In addition, some models of quantum gravity predict that the quantum sideband contributions (*iii*) and (*iv*) will occur at slightly different frequency than the thermal sideband contribution (*iv*), in which case they could in principle be resolved separately.[6]

A handful of experiments have measured both optomechanical sidebands in the quantum regime.[17,18,16,19,13] Here we describe an experiment which extends these measurements to a mechanical oscillator with substantially greater effective mass $m$ and lower frequency $\omega_m$ than in previous work. Increased $m$ and decreased $\omega_m$ are important for realizing, e.g., the proposals in Refs.[5,6]. More broadly, the particular type of oscillator used in this work (a $Si_3N_4$ membrane) has been shown to be well-suited to a range of other applications in quantum optomechanics.[8,9,20,21,22,23,24,25,26,27,28]

In the experiments described here, both sidebands are produced by a single laser and are measured simultaneously. This is in contrast with most earlier experiments, in which the sidebands were produced using two separate drives applied at once[16,19] or at different times.[17,13] We find that the behavior of the sidebands and other aspects of the data agree well with theory over a wide range of oscillator temperatures, extending to a mean phonon number $\bar{n} < 1$.

The experimental setup is shown in Fig. 1(b). The mechanical oscillator is a $Si_3N_4$ membrane with dimensions 1 mm × 1 mm × 50 nm.[29] The mechanical mode of interest is the membrane's (2,2) vibrational mode (in which the membrane is bisected by a nodal line along each in-plane axis).[30] This mode has effective mass $m$ = 43 ng, resonant frequency $\omega_m/2\pi$ = 705.2 kHz, and mechanical linewidth $\gamma_m/2\pi$ that varies between 0.07 – 0.14 Hz. The membrane is positioned inside a free-space Fabry-Pérot optical cavity with linewidth $\kappa/2\pi$ = 165 kHz (corresponding to finesse = 40,000). The ratio $\omega_m/\kappa$ = 4.3 is large enough to allow laser cooling to $\bar{n} < 1$[31,32] but small enough to ensure that the motional sidebands are not too strongly filtered by the cavity. The cavity is single-sided, with $\kappa_{in}/\kappa$ = 0.4. All measurements are performed in reflection.

The membrane and optical cavity are mounted in a $^3$He cryostat. Laser light is brought into the cryostat via an optical fiber and then coupled to the cavity via free-space optics mounted in the cryostat. Details of the cryogenic setup are given in Ref. [25].

Two lasers are used for the experiments described here: one for measurements and one for cooling (ML and CL in Fig. 1(b)). Both are continuous wave Nd:YAG lasers with wavelength $\lambda$ = 1,064 nm. The two lasers address cavity modes whose longitudinal mode numbers differ by two. To accomplish this, the CL is frequency-locked to the ML with an offset approximately equal to twice the cavity's free spectral range (2 × $\omega_{FSR}/2\pi \approx$ 8 GHz). The precise value of this offset is chosen so that the CL is detuned from its cavity mode by an amount $\Delta_{CL} \approx -\omega_m$ in order to provide optimal laser cooling.

Each laser passes through a filter cavity (FC1 and FC2 in Fig. 1(b)) with linewidth ~20 kHz, reducing classical noise power at 705 kHz by ~4,000. We do not observe any signature of classical laser noise in the measurements described here, consistent with independent characterization of the filtered beams.

The ML is split into two beams: a probe and a local oscillator (LO). An acousto-optic modulator (AOM) shifts the probe by $\omega_{LO}/2\pi$ = 80 MHz and an electro-optic modulator (EOM) provides phase modulation for Pound-Drever-Hall (PDH) locking to the cryogenic cavity. The probe beam, LO beam, and cooling beam are then combined into a single fiber and pass through a fiber-coupled AOM that is driven by a voltage controlled oscillator (VCO). The VCO frequency is tuned by the PDH error signal, so that all the beams simultaneously track fluctuations in the cryogenic cavity's resonant frequency. The spectrum of the beams and the cavity modes is illustrated in Fig. 1(c). The power in the probe and LO beam incident on the cavity were $P_{probe}$ = 32 μW and $P_{LO}$ = 1.57 mW.

As shown in Fig. 1(c), the probe beam drives the cavity nearly on resonance and acquires sidebands from the membrane's motion. Light from the cavity then falls on a photodiode, where the LO and motional sidebands mix to produce photocurrent signals at $\omega_{LO} \pm \omega_m$. Two demodulators, each with bandwidth 115 kHz, are used to simultaneously monitor the photocurrent at frequencies near $\omega_{LO} \pm \omega_m$. The power spectral density of these two records are $S_{II}^{(r)}(\omega)$ and $S_{II}^{(b)}(\omega)$ (corresponding to the vicinity of the red and blue sidebands), where $\omega$ is the frequency separation from the heterodyne carrier.

The noise floors of $S_{II}^{(r)}$ and $S_{II}^{(b)}$ were found to increase linearly with the total power incident on the photodiode, as expected for shot noise. The slope of this relationship was used to determine the overall gain $G$ of the detector. $G$ was found to differ by 0.5% between $\omega = \pm\omega_m$. The overall detection efficiency $\eta$ was determined by measuring the photodiode's detection efficiency and the loss in the beam path. The dark noise of the detector was found to differ by 1.5% between $\omega = \pm\omega_m$. The mechanical sidebands' central frequency ($\tilde{\omega}$) and linewidth ($\tilde{\gamma}$) differ from the membrane's intrinsic values ($\omega_m$ and $\gamma_m$) because of the optical spring and optical damping effects.[1] Measuring $\tilde{\omega}$ and $\tilde{\gamma}$ as a function of $P_{CL}$ (the power of the cooling laser) and fitting to the expected form provides the optomechanical coupling rate $g_0/2\pi = 2.2$ Hz. The detuning of the probe beam $\Delta_{probe}/2\pi = -6.5$ kHz was determined from $\tilde{\omega}$ and $\tilde{\gamma}$ when $P_{CL} = 0$ W. These values of $G$, $g_0$, $\eta$, and $\Delta_{probe}$ were used to convert the photocurrent spectra $S_{II}^{(r)}$ and $S_{II}^{(b)}$ into the spectra of the inferred membrane displacement $S_{xx}^{(r)}$ and $S_{xx}^{(b)}$, as described in Ref.[33].

Typical records of $S_{xx}^{(r)}$ and $S_{xx}^{(b)}$ are shown in Fig. 2 for different values of $P_{CL}$. Qualitatively, these measurements show the expected features described above: a noise floor corresponding to the detector's dark noise plus the shot noise, and motional sidebands at $\pm\omega_m$. As $P_{CL}$ increases, the noise floor increases (owing to the increased power incident on the photodiode) while the motional sidebands become smaller and broader (owing to optical cooling). At the higher values of $P_{CL}$, the area of the blue sideband is distinctly less than that of the red sideband.

In addition to these expected features, peaks are also visible at $\omega/2\pi = \pm 699$ kHz and $\pm 701$ kHz. These peaks appear symmetrically about $\omega_{LO}$ and are not observed when $\Delta_{probe} \gg \kappa$, indicating that they are due to motion in the cavity. The frequency, linewidth, and area of these peaks are independent of $P_{CL}$ and $\Delta_{CL}$, indicating that they are associated with mechanical modes that are too stiff to be influenced by radiation pressure. This suggests that these peaks are not due to the membrane, and are likely due to thermal motion of the cavity spacer and/or mirrors.

For each value of $P_{CL}$ the measured $S_{xx}^{(r)}$ and $S_{xx}^{(b)}$ were fit to the expected form[33]

$$S_{xx}^{(r,b)}(\omega) = b^{(r,b)} + s^{(r,b)}(\tilde{\gamma}/2)^2[(|\omega| - \tilde{\omega})^2 + (\tilde{\gamma}/2)^2]^{-1} \quad (1)$$

These fits are shown as black curves in Fig. 2. Six fitting parameters are used: $\tilde{\omega}$ (the center frequency of both sidebands), $\tilde{\gamma}$ (the linewidth of both sidebands), $b^{(r)}$ and $b^{(b)}$ (the noise floors of the red and blue sidebands), and $s^{(r)}$ and $s^{(b)}$ (the amplitudes of the red and blue sidebands). As described above, $b^{(r)}$ and $b^{(b)}$ scale linearly with $P_{CL}$, and are consistent with a gain difference 0.5% between the red and blue sidebands.

Figure 3 shows a summary of the fitting results as a function of $P_{CL}$. Figure 3(b) shows $\tilde{\gamma}/2\pi$, which increases with $P_{CL}$ and reaches a maximum value 4.86 ± 0.62 kHz. Figure 3(c) shows the inverse of the sidebands' areas $1/A^{(r)}$ and $1/A^{(b)}$ (where $A^{(r,b)} \equiv \frac{1}{4}\tilde{\gamma}s^{(r,b)}$). Both $1/A^{(r)}$ and $1/A^{(b)}$ increase with $P_{CL}$, but at higher values of $P_{CL}$, $1/A^{(r)}$ saturates while $1/A^{(b)}$ continues to increase. Figure 3(d) shows a measure of the sideband asymmetry, $\xi \equiv (A^{(r)}/A^{(b)} - 1)$. The asymmetry $\xi$ increases with $P_{CL}$, reaching a maximum value of 1.18 ± 0.32. Errors quoted in the text and error bars in the figures correspond to one standard deviation of statistical uncertainty in the fits to Eq. (1).

The membrane's mean phonon number $\bar{n}$ can be inferred from these measurements in a number of ways. Below, we use four different methods, each of which is directly connected to one of the quantities $\xi$, $A^{(r)}$, $A^{(b)}$, and $\tilde{\gamma}$.

*Sideband asymmetry*: As summarized above (and discussed in detail elsewhere[15,16]) the ratio of the sideband areas gives a direct estimate of the mean phonon number: $\bar{n} = 1/\xi$. This method has the advantage of being independent of the absolute calibration of the heterodyne signal and does not require knowledge of $\gamma_m$ or the bath temperature $T_{bath}$. It does require knowledge of $\Delta_{probe}$ (since a detuned probe beam results in unequal filtering of the sidebands by the cavity,[33] leading to an asymmetry that is independent of $\bar{n}$) and assumes that classical noise of the laser can be neglected. The values of $\bar{n}$ resulting from this method are shown as the green points in Fig. 3(e). The lowest value is $\bar{n} = 0.84 \pm 0.22$.

*Absolutely calibrated displacement*: Each of the calibrated displacement spectra $S_{xx}^{(r)}$ and $S_{xx}^{(b)}$ can be used to estimate $\bar{n}$ via the equipartition theorem.[33] For the spectrum from the blue sideband $\bar{n} = A^{(b)}/2x_{ZP}^2$, while for the spectrum from the red sideband $\bar{n} + 1 = A^{(r)}/2x_{ZP}^2$, where $x_{ZP} = (\hbar/2m\omega_m)^{1/2}$. These estimates do not require knowledge of $\gamma_m$ or $T_{bath}$, but do depend upon the absolute calibration of the heterodyne signal. The values of $\bar{n}$ resulting from this method are shown as the red and blue points in Fig. 3(e). The lowest value is $\bar{n} = 0.88 \pm 0.27$ (from the red sideband) and $\bar{n} = 0.86 \pm 0.16$ (from the blue sideband).

*Total damping rate*: When the probe beam and cooling beam address different cavity modes, the total damping rate of the mechanical oscillator can be used to estimate $\bar{n}$ via[31,32,33]

$$\bar{n} = (\bar{n}_{bath}\gamma_m + \bar{n}_{CL}\gamma_{CL} + \bar{n}_{probe}\gamma_{probe})/(\gamma_m + \gamma_{CL} + \gamma_{probe}) \qquad (2)$$

Here $\bar{n}_{bath} = k_B T_{bath}/\hbar\tilde{\omega}$ is the mean phonon number of an oscillator in equilibrium with the thermal bath,

and $\bar{n}_{\text{CL,probe}} = -\left((\omega_m + \Delta_{\text{CL,probe}})^2 + \left(\frac{\kappa}{2}\right)^2\right)/4\omega_m\Delta_{\text{CL,probe}}$ each represent the mean phonon number of an oscillator in equilibrium with the quantum fluctuations of a driven cavity mode. $\gamma_{\text{CL}}$ and $\gamma_{\text{probe}}$ are the optical damping introduced by the cooling and probe beams, and $\gamma_m + \gamma_{\text{CL}} + \gamma_{\text{probe}} = \tilde{\gamma}$.

This method is independent of the heterodyne calibration, but requires knowledge of $\gamma_m$ and $T_{\text{bath}}$. Mechanical ringdown measurements result in a value of $\gamma_m/2\pi$ that varied between 0.07 – 0.14 Hz; for the analysis presented here we use $\gamma_m/2\pi = 0.14$ Hz. To estimate $T_{\text{bath}}$, two $RuO_2$ thermometers were monitored during the experiment. One was attached to the $^3$He pot, while the other was attached to the stage on which the membrane chip was mounted. These two records, $T_{\text{pot}}$ and $T_{\text{stage}}$, are shown as open points in Fig. 3(a). Since neither thermometer was in direct contact with the membrane chip, we assume that $T_{\text{bath}}$ is a weighted average of these two readings: $T_{\text{bath}} = \alpha T_{\text{stage}} + (1-\alpha)T_{\text{pot}}$. We choose $\alpha$ to be the value for which the $\bar{n}(P_{\text{CL}})$ determined from Eq. (2) have the least squared difference from the $\bar{n}(P_{\text{CL}})$ determined from the sideband asymmetry (green points in Fig. 3(e)). This fitting procedure gives $\alpha = 0.498$. The corresponding $T_{\text{bath}}$ is shown as the solid points in Fig. 3(a). The values of $\bar{n}$ resulting from this method are the yellow points in Fig. 3(e). The lowest value is $\bar{n} = 0.88 \pm 0.10$.

The solid lines in Figs. 3(b)-(e) are the predicted values of $\tilde{\gamma}$, $1/A^{(r)}$, $1/A^{(b)}$, $\xi$, and $\bar{n}$. In each case they are calculated from the measured values of the parameters $\Delta_{\text{CL}}$, $\Delta_{\text{probe}}$, $P_{\text{CL}}$, $P_{\text{probe}}$, $P_{\text{LO}}$, $\gamma_m$, $\kappa$, $\kappa_{\text{in}}$, $T_{\text{bath}}$, $g_0$, $m_{\text{eff}}$, $\omega_m$, $\eta$, and $G$ using the expressions in Ref.[33].

The four estimates of the membrane's mean phonon number shown in Fig. 3(e) are based on different physical principles, and on different aspects of the data. The systematic and statistical uncertainties in these estimates are not completely independent, but the agreement between them over a wide range of temperature indicates that the system is accurately described by the standard theory of optomechanical systems in the quantum regime.

In the course of this work we became aware of parallel studies.[34]

We acknowledge support from AFOSR (FA9550-90-1-0484 and FA9550-15-1-0270) and NSF (DMR-1301798 and PHY-0855455). K. B. acknowledges financial support from The Research Council of Norway and from the Danish Council for Independent Research under the Sapere Aude program. We thank Yanbei Chen, Aashish Clerk, and Florian Marquardt for helpful discussions, and Huub Janssen and Yeubin Ning for technical assistance.

**Figure Captions**

Fig. 1: Schematic of the measurements. (a) Contributions to a heterodyne measurement of light leaving an optomechanical cavity, assuming the oscillator is at $T = 0$ and the cavity is driven on resonance by a single laser. Contributions from the shot noise (green), the oscillator's zero point motion (red), and the oscillator's response to the quantum fluctuations of the cavity field (blue) are shown in the vicinity of the red sideband (left) and the blue sideband (right). The total signal is the black curve. The vertical axis is the power spectral density of the photocurrent; the horizontal axis is measurement frequency. (b) The experimental setup. Free space beams are colored lines; optical fibers are hollow lines; electrical circuits are thick black lines. Two separate lasers (ML and CL) pass through filter cavities (FC1 and FC2). The probe beam is shifted by AOM1, while AOM2 tracks fluctuations in the cryogenic cavity. Light is delivered to (and collected from) the cryostat by a circulator. Control circuits, photodiodes, and fiber couplers are indicated by triangles, semicircles, and ovals, respectively. The mechanical oscillator is shown in purple. (c) The spectrum of the lasers (orange, light green, and dark green), the cavity modes (black), and the mechanical sidebands (red and blue).

Fig. 2: Motional sidebands of the mechanical oscillator. The membrane's displacement power spectral density $S_{xx}$ is plotted as a function of measurement frequency $\omega$. The red data (left panels) show $S_{xx}^{(r)}$, i.e., $S_{xx}$ near the red sideband. The blue data (right panels) show $S_{xx}^{(b)}$, i.e., $S_{xx}$ near the blue sideband. Each row corresponds to a different cooling laser power: from top to bottom, $P_{CL}$ = 0, 34, 158, 415 µW. The black line is the fit described in the text. The red/blue shading indicates the Lorentzian portion of the fit, specified by $\tilde{\omega}$, $\tilde{\gamma}$, $s^{(r)}$, and $s^{(b)}$. The gray shading indicates the fitted noise floor (shot noise plus dark noise), specified by $b^{(r)}$ and $b^{(b)}$. The detuning of the probe beam causes the displacement imprecision of the blue data to differ slightly from that of the red data; as a visual guide, the vertical axis of the blue data is shifted to compensate for this difference. The data was fit over the range 702 kHz $\leq |\omega/2\pi| \leq$ 714 kHz. As $P_{CL}$ increases, the sidebands broaden and shrink, owing to laser cooling. In the lowest panel, the ratio of their areas is $\xi + 1 = 2.18 \pm 0.32$, corresponding to $\bar{n} = 0.84 \pm 0.22$.

Fig. 3: Behavior as a function of cooling laser power $P_{CL}$. (a) Temperature recorded by two thermometers ($T_{pot}$ and $T_{stage}$, hollow points) and the membrane's inferred bath temperature $T_{bath}$ (solid points). Note $T_{pot} < T_{stage}$. (b) Mechanical linewidth $\tilde{\gamma}$. (c) Inverse area of each sideband, $1/A^{(r)}$ and $1/A^{(b)}$. (d) Asymmetry of the sideband areas $\xi$. (e) Inverse mean phonon number $1/\bar{n}$, determined from $\xi$ (green), $A^{(r)}$ (red), $A^{(b)}$ (blue), and $\tilde{\gamma}$ (yellow). Solid lines in (b) – (e) are calculated values, as described in the text. In (b) – (e), each inset shows a detailed view of the data for low $P_{CL}$.

**Figure 1**

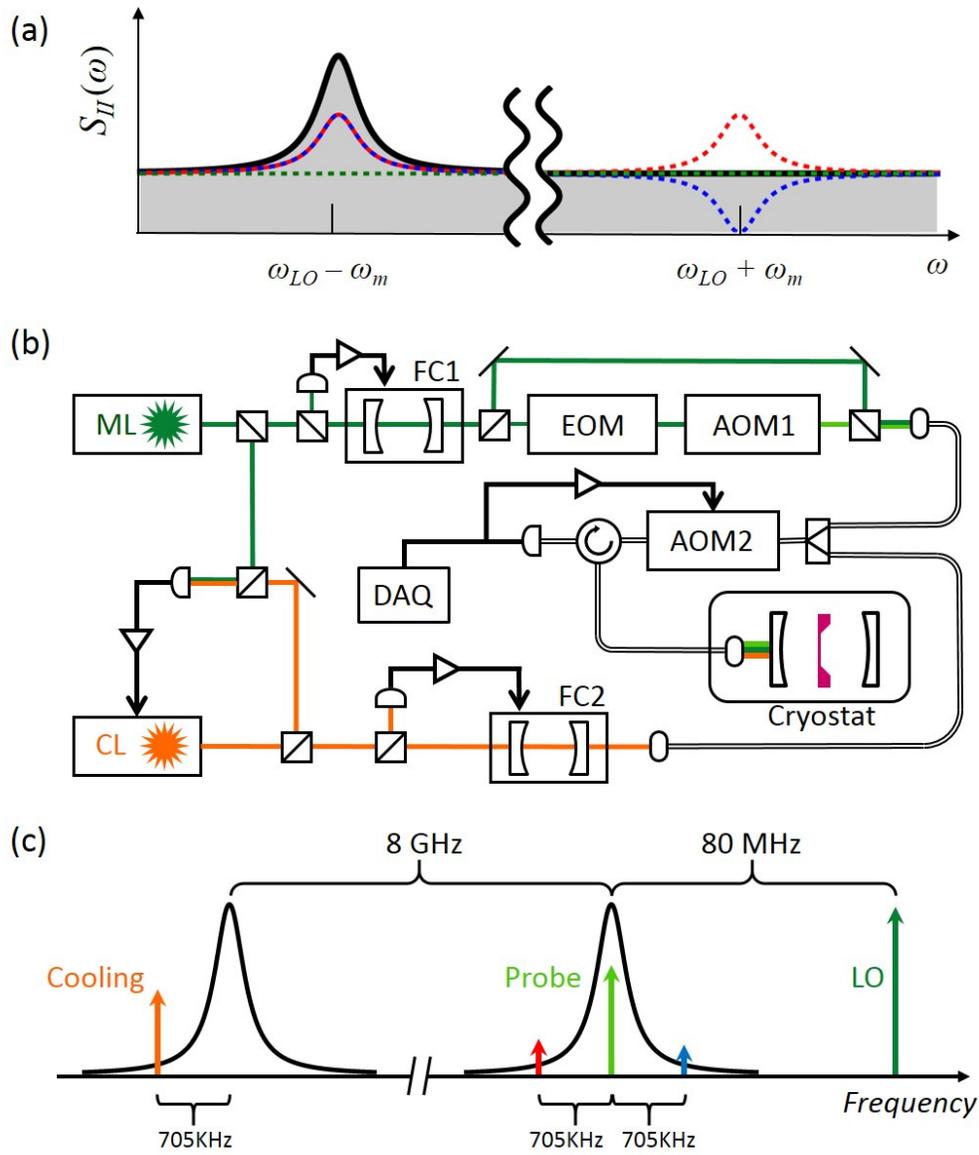

**Figure 2**

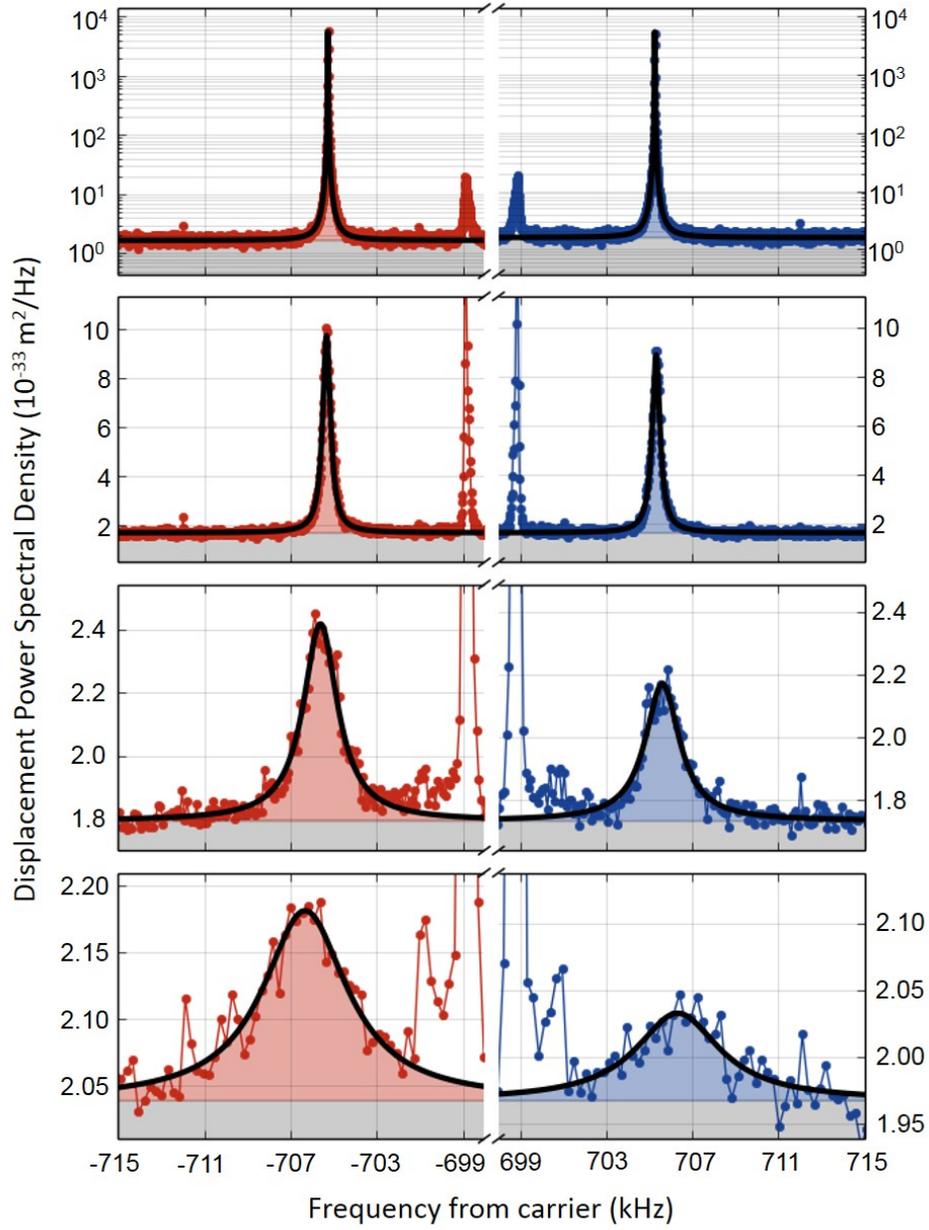

**Figure 3**

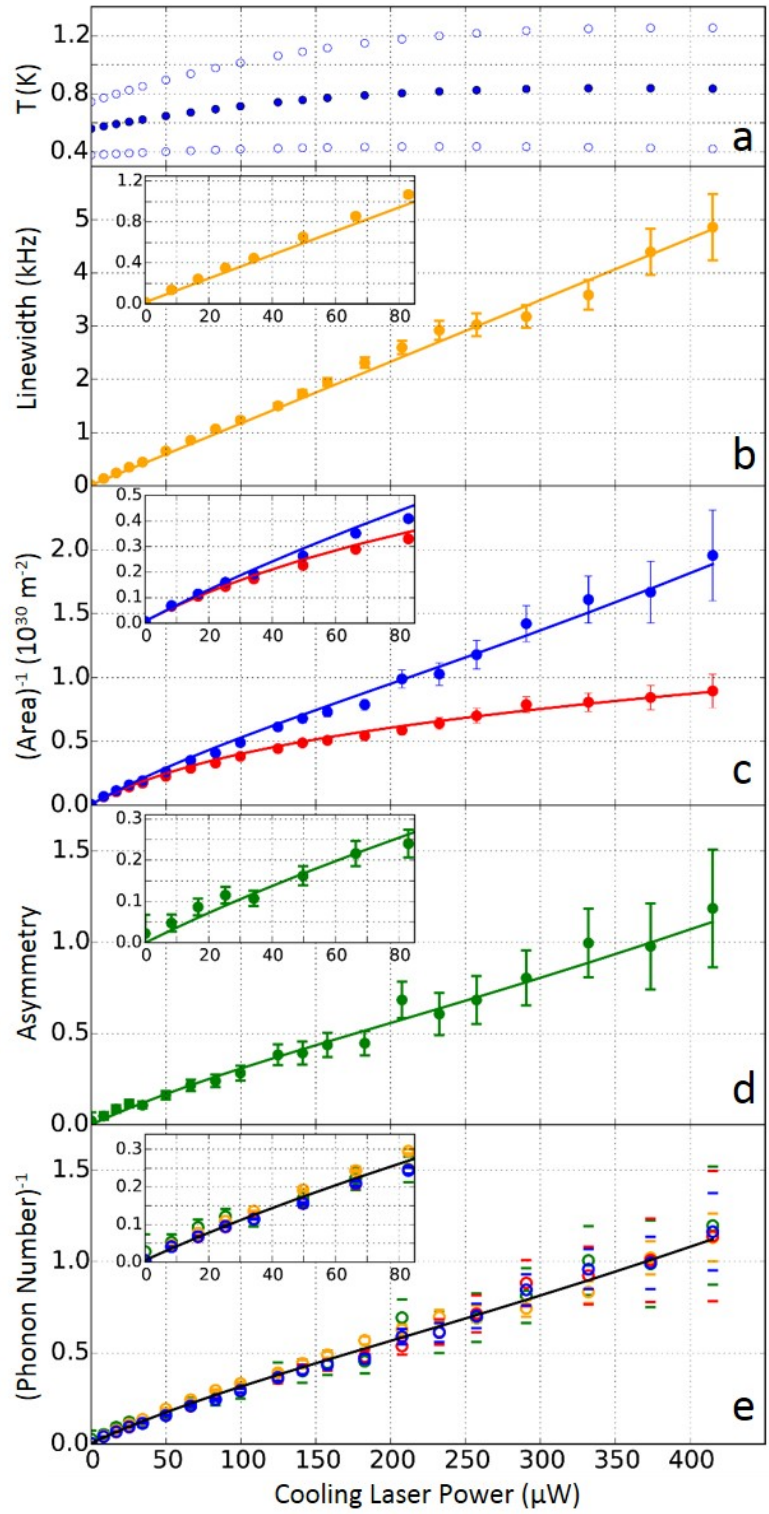